# A Nested Decomposition Method and Its Application for Coordinated Operation of Hierarchical Electrical Power Grids

Chenhui Lin
Department of Electrical Engineering, Tsinghua University, Beijing, 100084, China, linch11@yeah.net

Wenchuan Wu
Department of Electrical Engineering, Tsinghua University, Beijing, 100084, China, wuwench@tsinghua.edu.cn

Multilevel, multiarea, and hierarchically interconnected electrical power grids confront substantial challenges with the increasing integration of many volatile energy resources. The traditional isolated operation of interconnected power grids is uneconomical due to a lack of coordination; it may result in severe accidents that affect operational safety. However, the centralized operation of interconnected power grids is impractical, considering the operational independence and information privacy of each power grid. This paper proposes a nested decomposition method for the coordinated operation of hierarchical electrical power grids, which can achieve global optimization by iterating among upper- and lower-level power grids with exchange of boundary information alone. During each iteration, a projection function, which embodies the optimal objective value of a lower-level power grid projected onto its boundary variable space, is computed with second-order exactness. Thus, the proposed method can be applied widely to nonlinear continuous optimizations and can converge much more rapidly than existing decomposition methods. We conducted numerical tests of coordinated operation examples with a trilevel power grid that demonstrate the validity and performance of the proposed method.

*Key words*: nested decomposition method, coordinated operation, hierarchical electrical power grids, projection function
*History*: This paper was first submitted in June 2020.

## 1. Introduction

In real time, an electrical power control center monitors the local electrical power and operates the local power grid to enhance efficiency and prevent outages, which would create a security problem. Operational problems are essentially mathematical optimization problems, in which the grid-wide total generation cost is typically optimized to gain economic benefits, as well as to satisfy the operational and security constraints of that grid. In general, two types of operational problems are considered in electrical power grids: dynamic economic dispatch and optimal power flow. Dynamic economic dispatch jointly optimizes a number of future dispatch periods, through the use of an approximate, linearized, power-flow model. The generation cost functions of thermal generators are typically quadratic; thus, typical dynamic economic dispatch models involve quadratic programming. Optimal power flow only considers a single upcoming dispatch period; more precise, nonlinear, power-flow models are therefore used. Accordingly, the optimal power-flow model is typically a nonlinear programming problem.

Electrical power grids are hierarchically interconnected; thus, the operation, control, and management of each local power grid is hierarchical and independent. Conventionally, the boundaries of interconnected local power grids are determined and negotiated offline in an approximate manner. However, increasing numbers of uncertain and unstable volatile energy resources (e.g., wind turbines, photovoltaic panels, and electric vehicles) are undergoing integration into multi-level electrical power grids. The traditional manner of operating, which lacks coordination, may lead to unnecessary generation costs and severe security issues (e.g., load shedding) caused by insufficient local generation flexibility. Therefore, coordination is essential in the operation of multilevel hierarchical electrical power grids.

A straightforward coordinating method involves combining the operational models of separate local power grids and centrally optimizing the overall power grid, based on an integrated model. However, this is impractical considering the operational independence and information privacy of each power grid operator; these operators would prefer to avoid sharing their topology parameters or operational states with others for central optimization. Therefore, decomposition methods should be investigated to solve the integrated optimization model, while respecting each participant's privacy. Decomposition methods are typically implemented by iterating among constituent sub-problems. Each sub-problem is optimized locally; sub-problems then exchange limited boundary information with each other until the iteration converges.

Some research articles have discussed various types of decomposition methods and their applications to power grid operations. Benders decomposition (Benders 1962) can solve bilevel, coordinated, optimization problems; it was originally proposed to solve mixed-integer linear programming problems, in which the decomposition iterates between the complicated discrete component and the simple continuous component of the problem. The same iteration scheme can also be used to solve linear, continuous, optimization problems in electrical power grid operation (Li et al. 2015; Lin et al. 2016; Moya 2005; Sifuentes and Vargas 2007). Geoffrion (1972) extends the Benders decomposition method to generalized Benders decomposition, allowing the method to solve nonlinear programming problems. However, generalized Benders decomposition has poor convergence, which is a major barrier to its application in electrical power grids.

The original Benders decomposition, as well as its generalized form, can be classified as primal decomposition; these methods decompose the coordinated optimization problem directly by using boundary coupling variables. Another type of decomposition method, dual decomposition, exploits the dual nature of the original coordinated optimization problem (Conejo et al. 2006). Augmented Lagrangian decomposition (Hestenes 1969; Powell 1969) and the alternating direction method of multipliers (ADMM; Boyd et al. 2011) are two common implementations of dual decomposition. For typical applications of these methods in power grid operations, refer to Wang et al. 1995, Chen and Yang 2017, Li et al. 2016, and Zheng et al. 2015. The convergence speed of dual decomposition methods is typically uncertain. In extreme scenarios, the communication burden could be extremely heavy. Furthermore, existing works mostly focus on bilevel decomposition. The extension of bilevel decomposition method to multilevel decomposition method requires further analysis.

In this paper, we propose a novel decomposition method to achieve coordinated operation of bilevel electrical power grids. In this method, the lower, more local problems are formulated as optimization problems with both internal and boundary variables. Each lower problem is projected into a boundary variable space with second-order information comprising its objective function and constraints, thus forming projection function expansions. Expansions of lower problems are iterated with the upper, more global problem to converge on the globally optimal solution. Because second-order information of lower problems is considered in each iteration, the convergence speed of the proposed bilevel decomposition method is considerably more rapid than that of existing first-order-based methods (e.g., Benders decomposition). We then develop a nested, recursive framework to extend the proposed bilevel method into multilevel coordination. The convergence and optimality of the proposed, nested, decomposition method is proven theoretically by using convex optimization problems with continuously differentiable objectives and constraints. Finally, we conduct numerical simulations of the dynamic economic dispatch and optimal power flow of a trilevel electrical power grid to demonstrate the applicability and efficiency of the proposed method.

## 2. Formulation of the Hierarchical Coordinated Operation Problem

Electrical power grids are multilevel and hierarchically interconnected. Figure 1 shows a simple example of a trilevel power grid, which contains a transmission grid in the top level, two distribution grids in the middle level, and six microgrids in the bottom level. All sub-grids are connected as a tree structure, in which the boundary information of each local power grid can only be coupled with its single upper-grid level or multiple lower-level grids.

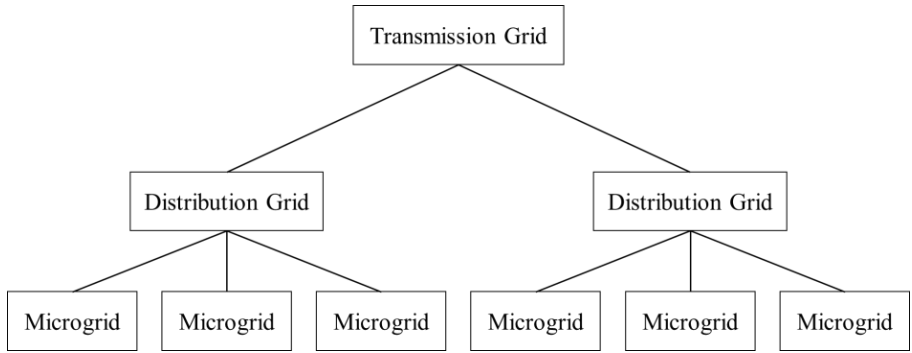

Figure 1. Example structure for a trilevel power grid.

Consider a local power grid in a particular level; its operational problem can be formulated as an optimization problem:

$$\min_{x_{m,n}, u_{m,n}, l_{m,n}} f_{m,n}\left(x_{m,n}\right) \\ \text{s.t. } G_{m,n}\left(x_{m,n}, u_{m,n}, l_{m,n}\right) \leq 0 \tag{1}$$

where subscripts $(\ )_{m,n}$ denote the local power grid in level, $m$, whose index number is $n$. Each of the following components applies to that grid: $x_{m,n}$ is the vector of internal variables; $u_{m,n}$ is the vector of upper boundary variables as it couples with the connected power grid at the level immediately above the current level; $l_{m,n}$ is the vector of lower boundary variables as it couples with the connected power grids at the level immediately below the current level; $f_{m,n}\left(x_{m,n}\right)$ is the objective function; and $G_{m,n}\left(x_{m,n}, u_{m,n}, l_{m,n}\right) \leq 0$ is the constraint vector.

Because power grids are interconnected, the boundary variables of adjacent, local, power grids should match each other. Accordingly, a general formulation describing the coordinated operation of multilevel hierarchical electrical power grids can be represented as follows:

$$\min_{x_{m,n},u_{m,n},l_{m,n}} \sum_{m \in M} \sum_{n \in N(m)} f_{m,n}(x_{m,n})$$

$$\text{s.t.} \quad G_{m,n}(x_{m,n},u_{m,n},l_{m,n}) \leq 0, \forall n \in N(m), \forall m \in M \quad (2)$$

$$u_{m,n} = I_{m,n} l_{m-1,U(m,n)}, \forall n \in N(m), \forall m \in M \setminus \{1\}$$

where $M$ is the index set of all levels; $N(m)$ is the index set of all local power grids in level $m$; $U(m,n)$ is the index number of the upper connected problem of the local power grid $n$ in level $m$; $I_{m,n}$ is the mapping matrix between the upper boundary variables $u_{m,n}$ and lower boundary variables $l_{m-1,U(m,n)}$; and $u_{m,n} = I_{m,n} l_{m-1,U(m,n)}$ is the boundary coupling constraint between the local power grid $n$ in level $m$ and its connected upper power grid.

## 3. Decomposition Method for Bilevel Coordination

We first consider the case in which there are only two electrical power grid levels in the coordinated operation problem. The upper level is denoted level 1; the lower level is denoted level 2. Correspondingly, $m$ in Equation (2) is either 1 or 2, and the coordinated operation problem of bilevel electrical power grids is written as follows:

$$\min_{x_1,x_{2,n},l_1,u_{2,n}} f_1(x_1) + \sum_{n \in N(2)} f_{2,n}(x_{2,n})$$

$$\text{s.t.} \quad G_1(x_1,l_1) \leq 0 \quad (3)$$

$$G_{2,n}(x_{2,n},u_{2,n}) \leq 0, \forall n \in N(2)$$

$$u_{2,n} = I_{2,n} l_1, \forall n \in N(2)$$

Note that, because the upper level has only one power grid, its corresponding subscript $(\ )_1$ neglects the intralevel power grid index number and only contains the level number.

In electrical power grid operation problems, the objectives and constraints might be linear or nonlinear, depending on the specific scenarios and models selected. We consider the case in which objectives and constraints are continuously differentiable and convex.

### 3.1 Problem Decomposition

The bilevel, coordinated, operation problem is decomposed into an upper problem and several lower problems. The upper problem optimizes its own variables:

$$\min_{x_1,l_1} f_1(x_1)$$

$$\text{s.t.} \quad G_1(x_1,l_1) \leq 0 \quad (4)$$

For each lower problem, its optimization also must satisfy boundary coupling constraints. The formulation of the lower problem for $n$ is as follows:

$$\min_{x_{2,n},u_{2,n}} f_{2,n}(x_{2,n})$$

$$\text{s.t.} \quad G_{2,n}(x_{2,n},u_{2,n}) \leq 0 \quad (5)$$

$$u_{2,n} = I_{2,n} \tilde{l}_1$$

where $\tilde{l}_1$ is the boundary variable solution from the upper problem in Equation (4).

In the lower problem (Equation (5)), $I_{2,n}$ is a constant mapping matrix from $\tilde{l}_1$ to $u_{2,n}$, and $I_{2,n}\tilde{l}_1$ is considered a vector of fixed parameters. If this boundary parameter vector, $I_{2,n}\tilde{l}_1$, is chosen improperly, the lower problem might be unsolvable. In this case, a relaxed corresponding problem can be formed, which is always solvable:

$$\min_{x_{2,n}, u_{2,n}} f_{2,n}(x_{2,n}) + \left(c_{2,n}^{pty}\right)^T \left|u_{2,n} - I_{2,n}\tilde{l}_1\right|$$
$$s.t. \ G_{2,n}(x_{2,n}, u_{2,n}) \leq 0 \tag{6}$$

where $c_{2,n}^{pty}$ is a positive penalty parameter vector of the relaxation term. If this vector is sufficiently large, relaxation will not introduce errors to the original global problem in Equation (3).

Because the relaxed problem in Equation (6) can also be rewritten in its compact form in Equation (5), we continue to use the latter formulation as the lower problem in the following discussions and consider it to be solvable, regardless of the chosen $I_{2,n}\tilde{l}_1$.

**3.2 Lower Problem Projection**

Considering that $I_{2,n}\tilde{l}_1$ is a parameter vector in the lower problem (Equation (5)), its value affects the optimal solution and optimal objective value of that equation. Thus, a map from $I_{2,n}\tilde{l}_1$ to the optimal objective value of Equation (5) can be established, which is denoted by the function $J_{2,n}(I_{2,n}\tilde{l}_1)$. This mapping function can also be described as a projection of the lower problem to the boundary parameter vector space. It is defined mathematically as follows:

$$J_{2,n}(I_{2,n}\tilde{l}_1) = \min_{x_{2,n}, u_{2,n}} f_{2,n}(x_{2,n})$$
$$s.t. \ G_{2,n}(x_{2,n}, u_{2,n}) \leq 0 \tag{7}$$
$$u_{2,n} = I_{2,n}\tilde{l}_1$$

The above problem can be further simplified as follows:

$$J_{2,n}(I_{2,n}\tilde{l}_1) = \min_{x_{2,n}} f_{2,n}(x_{2,n})$$
$$s.t. \ G_{2,n}(x_{2,n}, I_{2,n}\tilde{l}_1) \leq 0 \tag{8}$$

Its optimal solution is denoted $\hat{x}_{2,n}$ and its dual multiplier at the optimal solution is denoted $\hat{\lambda}$. The Lagrange function of Equation (8) is as follows:

$$L(x_{2,n}, \lambda) = f_{2,n}(x_{2,n}) + \lambda^T G_{2,n}(x_{2,n}, I_{2,n}\tilde{l}_1) \tag{9}$$

At the optimal solution, the Karush–Kuhn–Tucker conditions give the following:

$$\left[\frac{df_{2,n}}{dx_{2,n}}\right]^T + \left[\frac{\partial G_{2,n}}{\partial x_{2,n}}\right]^T \hat{\lambda} = 0 \tag{10}$$

$$diag(\hat{\lambda}) G_{2,n}(\hat{x}_{2,n}, I_{2,n}\tilde{l}_1) = 0$$

where $diag(\hat{\lambda})$ denotes the diagonal matrix constructed from vector $\hat{\lambda}$.

Next, we consider the relationship between the boundary parameter vector and the optimal solution of Equation (8). For any $I_{2,n}l_1$ around $I_{2,n}\tilde{l}_1$, the corresponding optimal primal and dual solutions are functions of $I_{2,n}l_1$, which are denoted by $x_{2,n}(I_{2,n}l_1)$ and $\lambda(I_{2,n}l_1)$, respectively. As $I_{2,n}l_1$ changes, the above Karush–Kuhn–Tucker conditions remain applicable. Therefore,

$$\frac{d\left(\left[\frac{df_{2,n}}{dx_{2,n}}\right]^T + \left[\frac{\partial G_{2,n}}{\partial x_{2,n}}\right]^T \lambda\right)}{dI_{2,n}l_1} = 0 \tag{11}$$

$$\frac{d(diag(\lambda) G_{2,n}(x_{2,n}, I_{2,n}l_1))}{dI_{2,n}l_1} = 0$$

Expanding the two equations above gives the following:

$$\left[\frac{d^2 f_{2,n}}{dx_{2,n}^2}\right]\left[\frac{dx_{2,n}}{dI_{2,n}l_1}\right] + \left[\frac{\partial G_{2,n}}{\partial x_{2,n}}\right]^T\left[\frac{d\lambda}{dI_{2,n}l_1}\right] + \left(\left[\frac{\partial^2(\hat{\lambda}^T G_{2,n})}{\partial x_{2,n}^2}\right]\left[\frac{dx_{2,n}}{dI_{2,n}l_1}\right] + \left[\frac{\partial^2(\hat{\lambda}^T G_{2,n})}{\partial x_{2,n}\partial I_{2,n}l_1}\right]\right) = 0 \tag{12}$$

$$diag(\hat{\lambda})\left(\left[\frac{\partial G_{2,n}}{\partial x_{2,n}}\right]\left[\frac{dx_{2,n}}{dI_{2,n}l_1}\right] + \left[\frac{\partial G_{2,n}}{\partial I_{2,n}l_1}\right]\right) + diag(G_{2,n}(\hat{x}_{2,n}, I_{2,n}\tilde{l}_1))\left[\frac{d\lambda}{dI_{2,n}l_1}\right] = 0 \tag{13}$$

The first-order derivatives of the optimal primal and dual solutions, with respect to $I_{2,n}l_1$, can be solved as follows:

$$\begin{bmatrix}\dfrac{dx_{2,n}}{dI_{2,n}l_1} \\ \dfrac{d\lambda}{dI_{2,n}l_1}\end{bmatrix} = -\begin{bmatrix}\left[\dfrac{d^2 f_{2,n}}{dx_{2,n}^2}\right] + \left[\dfrac{\partial^2(\hat{\lambda}^T G_{2,n})}{\partial x_{2,n}^2}\right] & \left[\dfrac{\partial G_{2,n}}{\partial x_{2,n}}\right]^T \\ diag(\hat{\lambda})\left[\dfrac{\partial G_{2,n}}{\partial x_{2,n}}\right] & diag(G_{2,n}(\hat{x}_{2,n}, I_{2,n}\tilde{l}_1))\end{bmatrix}^{-1}\begin{bmatrix}\left[\dfrac{\partial^2(\hat{\lambda}^T G_{2,n})}{\partial x_{2,n}\partial I_{2,n}l_1}\right] \\ diag(\hat{\lambda})\left[\dfrac{\partial G_{2,n}}{\partial I_{2,n}l_1}\right]\end{bmatrix} \tag{14}$$

For simplicity, the above first-order derivatives are denoted as follows:

$$\frac{dx_{2,n}}{dI_{2,n}l_1} = R \tag{15}$$

$$\frac{d\lambda}{dI_{2,n}l_1} = S \tag{16}$$

A concrete form of the projection function in Equation (7) cannot be directly computed. However, its second-order Taylor series expansion can be computed, based on the first-order derivatives from Equation (14).

For any $I_{2,n}l_1$, $x_{2,n}(I_{2,n}l_1)$ and $\lambda(I_{2,n}l_1)$ are related functions. The projection function, $J_{2,n}(I_{2,n}l_1)$, is equal to the value of the Lagrange function at its optimal value:

$$J_{2,n}(I_{2,n}l_1) = L(x_{2,n}(I_{2,n}l_1), \lambda(I_{2,n}l_1), I_{2,n}l_1) = f_{2,n}(x_{2,n}) + \lambda^T G_{2,n}(x_{2,n}, I_{2,n}l_1) \tag{17}$$

The second-order Taylor series expansion of $J_{2,n}(I_{2,n}l_1)$ is as follows:

$$J_{2,n}(I_{2,n}\tilde{l}_1) + \left[\frac{dJ_{2,n}}{dI_{2,n}l_1}\right](I_{2,n}l_1 - I_{2,n}\tilde{l}_1) + \frac{1}{2}(I_{2,n}l_1 - I_{2,n}\tilde{l}_1)^T \left[\frac{d^2 J_{2,n}}{d(I_{2,n}l_1)^2}\right](I_{2,n}l_1 - I_{2,n}\tilde{l}_1) \tag{18}$$

The first-order derivative of $J_{2,n}(I_{2,n}l_1)$, with respect to $I_{2,n}l_1$, is as follows:

$$\frac{dJ_{2,n}}{dI_{2,n}l_1} = \left[\frac{\partial L}{\partial x_{2,n}}\right]\left[\frac{dx_{2,n}}{dI_{2,n}l_1}\right] + \left[\frac{\partial L}{\partial \lambda}\right]\left[\frac{d\lambda}{dI_{2,n}l_1}\right] + \left[\frac{\partial L}{\partial I_{2,n}l_1}\right] = \frac{\partial L}{\partial I_{2,n}l_1} = \left[\frac{\partial(\hat{\lambda}^T G_{2,n})}{\partial I_{2,n}l_1}\right] \tag{19}$$

This simplification is based on the observation that the optimal solutions are saddle points of Lagrange functions.

The second-order derivative of $J_{2,n}(I_{2,n}l_1)$, with respect to $I_{2,n}l_1$, is as follows:

$$\frac{d^2 J_{2,n}}{d(I_{2,n}l_1)^2} = \frac{d\left[\frac{dJ_{2,n}}{dI_{2,n}l_1}\right]^T}{dI_{2,n}l_1} = \frac{d\left(\left[\frac{dx_{2,n}}{dI_{2,n}l_1}\right]^T\left[\frac{\partial L}{\partial x_{2,n}}\right]^T + \left[\frac{d\lambda}{dI_{2,n}l_1}\right]^T\left[\frac{\partial L}{\partial \lambda}\right]^T + \left[\frac{\partial L}{\partial I_{2,n}l_1}\right]^T\right)}{dI_{2,n}l_1} \tag{20}$$

The above equation can be split into three terms, of which the first is as follows:

$$\frac{d\left(\left[\frac{dx_{2,n}}{dI_{2,n}l_1}\right]^T\left[\frac{\partial L}{\partial x_{2,n}}\right]^T\right)}{dI_{2,n}l_1} = \left[\frac{\partial L}{\partial x_{2,n}}\right]^T\left[\frac{d^2 x_{2,n}}{d(I_{2,n}l_1)^2}\right] + \left[\frac{dx_{2,n}}{dI_{2,n}l_1}\right]^T \frac{d\left[\frac{\partial L}{\partial x_{2,n}}\right]^T}{dI_{2,n}l_1}$$

$$= \left[\frac{dx_{2,n}}{dI_{2,n}l_1}\right]^T \left(\left[\frac{\partial^2 L}{\partial x_{2,n}^2}\right]\left[\frac{dx_{2,n}}{dI_{2,n}l_1}\right] + \left[\frac{\partial^2 L}{\partial x_{2,n}\partial \lambda}\right]\left[\frac{d\lambda}{dI_{2,n}l_1}\right] + \left[\frac{\partial^2 L}{\partial x_{2,n}\partial I_{2,n}l_1}\right]\right) \tag{21}$$

The second term is as follows:

$$\frac{d\left(\left[\frac{d\lambda}{dI_{2,n}l_1}\right]^T \left[\frac{\partial L}{\partial \lambda}\right]^T\right)}{dI_{2,n}l_1} = \left[\frac{\partial L}{\partial \lambda}\right]^T \left[\frac{d^2\lambda}{d(I_{2,n}l_1)^2}\right] + \left[\frac{d\lambda}{dI_{2,n}l_1}\right]^T \frac{d\left[\frac{\partial L}{\partial \lambda}\right]^T}{dI_{2,n}l_1}$$

$$= \left[\frac{d\lambda}{dI_{2,n}l_1}\right]^T \left(\left[\frac{\partial^2 L}{\partial \lambda \partial x_{2,n}}\right]\left[\frac{dx_{2,n}}{dI_{2,n}l_1}\right] + \left[\frac{\partial^2 L}{\partial \lambda^2}\right]\left[\frac{d\lambda}{dI_{2,n}l_1}\right] + \left[\frac{\partial^2 L}{\partial \lambda \partial I_{2,n}l_1}\right]\right) \quad (22)$$

The third term is as follows:

$$\frac{d\left(\left[\frac{\partial L}{\partial I_{2,n}l_1}\right]^T\right)}{dI_{2,n}l_1} = \left(\left[\frac{\partial^2 L}{\partial I_{2,n}l_1 \partial x_{2,n}}\right]\left[\frac{dx_{2,n}}{dI_{2,n}l_1}\right] + \left[\frac{\partial^2 L}{\partial I_{2,n}l_1 \partial \lambda}\right]\left[\frac{d\lambda}{dI_{2,n}l_1}\right] + \left[\frac{\partial^2 L}{\partial (I_{2,n}l_1)^2}\right]\right) \quad (23)$$

The second-order derivative of $J_{2,n}(I_{2,n}l_1)$ is the following sum of the above three terms:

$$\frac{d^2 J_{2,n}}{d(I_{2,n}l_1)^2} = \begin{bmatrix} R \\ S \\ I \end{bmatrix}^T \left[\frac{\partial^2 L}{\partial [x_{2,n} \;\; \lambda \;\; I_{2,n}l_1]^2}\right] \begin{bmatrix} R \\ S \\ I \end{bmatrix} \quad (24)$$

where $I$ is the identity matrix.

Note that the second-order derivative $\frac{\partial^2 L}{\partial \lambda^2}$ is 0. Therefore, the above equation can be further simplified as follows:

$$\frac{d^2 J_{2,n}}{d(I_{2,n}l_1)^2}$$

$$= \begin{bmatrix} R \\ I \end{bmatrix}^T \left[\frac{\partial^2 L}{\partial [x_{2,n} \;\; I_{2,n}l_1]^2}\right] \begin{bmatrix} R \\ I \end{bmatrix} + 2S^T \left(\left[\frac{\partial^2 L}{\partial \lambda \partial x_{2,n}}\right] R + \left[\frac{\partial^2 L}{\partial \lambda \partial I_{2,n}l_1}\right]\right)$$

$$= \begin{bmatrix} R \\ I \end{bmatrix}^T \left[\frac{\partial^2 L}{\partial [x_{2,n} \;\; I_{2,n}l_1]^2}\right] \begin{bmatrix} R \\ I \end{bmatrix} + 2S^T \left(\left[\frac{\partial G_{2,n}}{\partial x_{2,n}}\right] R + \left[\frac{\partial G_{2,n}}{\partial I_{2,n}l_1}\right]\right) \quad (25)$$

$$= \begin{bmatrix} R \\ I \end{bmatrix}^T \left[\frac{\partial^2 L}{\partial [x_{2,n} \;\; I_{2,n}l_1]^2}\right] \begin{bmatrix} R \\ I \end{bmatrix} + 2\left[\frac{d\lambda}{dI_{2,n}l_1}\right]^T \left(\left[\frac{\partial G_{2,n}}{\partial x_{2,n}}\right]\left[\frac{dx_{2,n}}{dI_{2,n}l_1}\right] + \left[\frac{\partial G_{2,n}}{\partial I_{2,n}l_1}\right]\right)$$

In the optimal solution, the constraints, $G_{2,n}(x_{2,n}, I_{2,n}\tilde{l}_1) \leq 0$, can be classified into active and inactive constraints, which are denoted by subscripts $[\;]_A$ and $[\;]_I$, respectively. When $I_{2,n}l_1$ changes, the optimal solution changes in a corresponding manner; however, the active or inactive nature of each constraint remains consistent, provided that the change of $I_{2,n}l_1$ is sufficiently small. Therefore, for active constraints, the values of $[G_{2,n}]_A$ and its derivatives, with respect to $I_{2,n}l_1$, are 0:

$$[G_{2,n}]_A = 0, \left[\frac{d[G_{2,n}]_A}{dI_{2,n}l_1}\right] = \left[\frac{\partial[G_{2,n}]_A}{\partial x_{2,n}}\right]\left[\frac{dx_{2,n}}{dI_{2,n}l_1}\right] + \left[\frac{\partial[G_{2,n}]_A}{\partial I_{2,n}l_1}\right] = 0 \tag{26}$$

Similarly, for inactive constraints, their corresponding dual variables and derivatives are 0:

$$[\lambda]_I = 0, \left[\frac{d[\lambda]_I}{dI_{2,n}l_1}\right] = 0 \tag{27}$$

Equation (25) can be further simplified to the following:

$$\begin{aligned}
\frac{d^2 J_{2,n}}{d(I_{2,n}l_1)^2} &= \begin{bmatrix} R \\ I \end{bmatrix}^T \left[\frac{\partial^2 L}{\partial [x_{2,n} \quad I_{2,n}l_1]^2}\right]\begin{bmatrix} R \\ I \end{bmatrix} + 2\left[\frac{d\lambda}{dI_{2,n}l_1}\right]^T \left(\left[\frac{\partial G_{2,n}}{\partial x_{2,n}}\right]\left[\frac{dx_{2,n}}{dI_{2,n}l_1}\right] + \left[\frac{\partial G_{2,n}}{\partial I_{2,n}l_1}\right]\right) \\
&= \begin{bmatrix} R \\ I \end{bmatrix}^T \left[\frac{\partial^2 L}{\partial [x_{2,n} \quad I_{2,n}l_1]^2}\right]\begin{bmatrix} R \\ I \end{bmatrix} + 2\left[\frac{d[\lambda]_A}{dI_{2,n}l_1}\right]^T \left(\left[\frac{\partial[G_{2,n}]_A}{\partial x_{2,n}}\right]\left[\frac{dx_{2,n}}{dI_{2,n}l_1}\right] + \left[\frac{\partial[G_{2,n}]_A}{\partial I_{2,n}l_1}\right]\right) \\
&\quad + 2\left[\frac{d[\lambda]_I}{dI_{2,n}l_1}\right]^T \left(\left[\frac{\partial[G_{2,n}]_I}{\partial x_{2,n}}\right]\left[\frac{dx_{2,n}}{dI_{2,n}l_1}\right] + \left[\frac{\partial[G_{2,n}]_I}{\partial I_{2,n}l_1}\right]\right) \\
&= \begin{bmatrix} R \\ I \end{bmatrix}^T \left[\frac{\partial^2 L}{\partial [x_{2,n} \quad I_{2,n}l_1]^2}\right]\begin{bmatrix} R \\ I \end{bmatrix}
\end{aligned} \tag{28}$$

Combining Equations (18), (19), and (28), the second-order Taylor expansion (denoted $JS_{2,n}(I_{2,n}l_1)$) of $J_{2,n}(I_{2,n}l_1)$ is as follows:

$$\begin{aligned}
JS_{2,n}(I_{2,n}l_1) &= f_{2,n}(\hat{x}_{2,n}) + \hat{\lambda}^T \left[\frac{\partial G_{2,n}}{\partial I_{2,n}l_1}\right](I_{2,n}l_1 - I_{2,n}\tilde{l}_1) \\
&\quad + \frac{1}{2}(I_{2,n}l_1 - I_{2,n}\tilde{l}_1)^T \begin{bmatrix} R_1 \\ I \end{bmatrix}^T \left[\frac{\partial^2 L}{\partial [x_{2,n} \quad I_{2,n}l_1]^2}\right]\begin{bmatrix} R_1 \\ I \end{bmatrix}(I_{2,n}l_1 - I_{2,n}\tilde{l}_1)
\end{aligned} \tag{29}$$

which can be regarded as an approximation of $J_{2,n}(I_{2,n}l_1)$.

If the lower problem is convex (with boundary parameters considered as variables), the projection function is also convex. In this case, the first-order Taylor expansion (denoted $JF_{2,n}(I_{2,n}l_1)$) of $J_{2,n}(I_{2,n}l_1)$ is one of its lower bounds, expressed as follows:

$$JF_{2,n}(I_{2,n}l_1) = f_{2,n}(\hat{x}_{2,n}) + \hat{\lambda}^T \left[\frac{\partial G_{2,n}}{\partial I_{2,n}l_1}\right](I_{2,n}l_1 - I_{2,n}\tilde{l}_1) \tag{30}$$

**3.3 Upper Problem Coordination**

Theoretically, the globally optimal solution can be solved by coordinating (at the upper level) all projection functions of lower problems:

$$\min_{x_1,l_1} f_1(x_1) + \sum_{n \in N(2)} J_{2,n}(I_{2,n}l_1) \tag{31}$$
$$s.t. \ G_1(x_1,l_1) \leq 0$$

However, the above coordination cannot be executed for two reasons. First, as analyzed in the previous sub-section, the projection function of each lower problem cannot be solved analytically. Second, the projection functions are typically not continuously differentiable, which causes difficult in solving the coordination problem in Equation (31).

Because approximations of these projection functions can be solved, an iteration procedure between the upper and lower problems is proposed to solve the globally optimal solution. The detailed procedures are as follows.

**Step 1:** The iteration number, $k$, is initialized as 1. The upper problem has the values of its boundary variables, $l_1^{(k)}$, initialized by solving:

$$\min_{x_1,l_1} f_1(x_1) \tag{32}$$
$$s.t. \ G_1(x_1,l_1) \leq 0$$

**Step 2:** Each lower problem receives its boundary parameter value $I_{2,n}l_1^{(k)}$ from the upper problem, and is optimized with that value:

$$\min_{x_{2,n}} f_{2,n}(x_{2,n}) \tag{33}$$
$$s.t. \ G_{2,n}\left(x_{2,n}, I_{2,n}l_1^{(k)}\right) \leq 0$$

**Step 3:** The first- and second-order Taylor series expansions of the projection function for each lower problem, denoted $JF_{2,n}^{(k)}(I_{2,n}l_1)$ and $JS_{2,n}^{(k)}(I_{2,n}l_1)$, respectively, are computed. The generated projection function expansions are sent to the upper problem.

**Step 4:** Optimization of the following equation allows coordination of projection function expansions from all lower problems by means of the upper problem:

$$\min_{x_1,l_1,o_{2,n}} f_1(x_1) + \sum_{n \in N(2)} o_{2,n}$$
$$s.t. \ G_1(x_1,l_1) \leq 0$$
$$o_{2,n} \geq JS_{2,n}^{(k)}(I_{2,n}l_1), \forall n \in N(2) \tag{34}$$
$$o_{2,n} \geq JF_{2,n}^{(i)}(I_{2,n}l_1), \forall i = 1,2,...,k-1, \forall n \in N(2)$$

where $o_{2,n}$ is an intermediate variable that reflects the objective value of the lower problem for $n$.

$k$ is iterated by 1. The optimal solutions of Equation (34) are denoted $x_1^{(k)}$, $l_1^{(k)}$, and $o_{2,n}^{(k)}$.

**Step 5:** Convergence is checked. The convergence criterion is $\left\| l_1^{(k)} - l_1^{(k-1)} \right\|_2 \leq \varepsilon$, where $\varepsilon$ is a small positive threshold. If convergence is observed, the solution from the latest iteration is the globally optimal solution; otherwise, the upper problem sends the updated $I_{2,n} l_1^{(k)}$ to each of the lower problems and the process returns to step 2.

In the above iteration procedure, the upper problem sends boundary values to each lower problem. For each lower problem, the first- and second-order Taylor series expansions of the projection function are computed for the given boundary value and sent back to the upper problem. In the upper coordination problem in Equation (34), only the second-order Taylor series expansions of the latest iteration are used, while all first-order Taylor series expansions of previous iterations are preserved. This is because the first-order expansions are lower bounds of the projection function in these lower problems, which are convex; the iteration procedure will thus converge when provided only with lower bounds at each iteration. The second-order expansions increase the iteration speed, as they better approximate the original projection function.

## 4. Nested Decomposition Method for Multilevel Hierarchical Coordination

The coordination of a multilevel, hierarchical, optimization problem can be considered a nested combination of several bilevel optimization problems. Specifically, each multilevel problem can be first decomposed into a top-level problem and several lower-level problems, where each lower-level problem is composed of a second-level problem and all much-lower-level problems covered by that second-level problem. It is straightforward to prove that each lower-level problem remains a multilevel optimization problem that can be further decomposed recursively until the problems in the bottom two levels have been decomposed. Thus, multilevel optimization can be decomposed in a nested manner.

Note that in this nested decomposition method, the projection procedure of the lower problem of each bilevel coordination must be performed in a decomposed manner (except in the recursive end level). Therefore, the main challenge of extending the bilevel decomposition method to multilevel decomposition is the computation of the first- and second-order Taylor series expansions of projection functions in a decomposed manner.

We assume that the total number of levels is $m_0$ (i.e., the set $M$ in Equation (2) is $\{1, 2, ..., m_0\}$). We therefore consider the problem number, $n_0$, in level, $m_0 - 1$, together with all lower problems included in $m_0$:

$$\min_{x_{m_0-1,n_0}, x_{m_0,n}, u_{m_0-1,n_0}, l_{m_0-1,n_0}, u_{m_0,n}} f_{m_0-1,n_0}\left(x_{m_0-1,n_0}\right) + \sum_{n \in L(m_0-1,n_0)} f_{m_0,n}\left(x_{m_0,n}\right)$$

$$\begin{aligned}
\text{s.t. } & G_{m_0-1,n_0}\left(x_{m_0-1,n_0}, u_{m_0-1,n_0}, l_{m_0-1,n_0}\right) \leq 0 \\
& u_{m_0-1,n_0} = I_{m_0-1,n_0} \tilde{l}_{m_0-2, U(m_0-1,n_0)} \\
& G_{m_0,n}\left(x_{m_0,n}, u_{m_0,n}\right) \leq 0, \forall n \in L(m_0-1,n_0) \\
& u_{m_0,n} = I_{m_0,n} l_{m_0-1,n_0}, \forall n \in L(m_0-1,n_0)
\end{aligned} \quad (35)$$

where $L(m_0-1, n_0)$ is the set of all problems in level $m_0$ covered by $n_0$ in $m_0-1$.

Our goal is the computation of Taylor series expansions of the projection functions at the given boundary value, $I_{m_0-1,n_0}\tilde{l}_{m_0-2,U(m_0-1,n_0)}$, from its connected upper problem in level $m_0-2$; that computation process must be decomposed.

The problem in Equation (35) is essentially a bilevel optimization problem between levels $m_0-1$ and $m_0$. Therefore, its optimal solution can be computed by iterating between those two levels with the method in Section 3.

During the iteration procedure for obtaining the optimal solution to Equation (35), the last upper coordination problem before convergence can be written as follows:

$$\min_{x_{m_0-1,n_0}, l_{m_0-1,n_0}, o_{m_0,n}} f_{m_0-1,n_0}\left(x_{m_0-1,n_0}\right) + \sum_{n \in L(m_0-1,n_0)} o_{m_0,n}$$

$$\begin{aligned}
\text{s.t. } & G_{m_0-1,n_0}\left(x_{m_0-1,n_0}, I_{m_0-2,n_0}\tilde{l}_{m_0-2,U(m_0-1,n_0)}, l_{m_0-1,n_0}\right) \leq 0 \\
& o_{m_0,n} \geq JS_{m_0,n}^{(k)}\left(I_{m_0,n} l_{m_0-1,n_0}\right), \forall n \in L(m_0-1,n_0) \\
& o_{m_0,n} \geq JF_{m_0,n}^{(i)}\left(I_{m_0,n} l_{m_0-1,n_0}\right), \forall i = 1, 2, \ldots, k-1, \forall n \in L(m_0-1,n_0)
\end{aligned} \quad (36)$$

This formulation only contains variables from $n_0$ in $m_0-1$. Therefore, it can be rewritten in a compact manner similar to the lower problem in Equation (8) for bilevel coordination; the Taylor series expansions of its projection functions can be computed. Recall that the entire computation process for projection expansions only requires second-order information regarding the lower-level problem. Indeed, the function $JS_{m_0,n}^{(k)}\left(I_{m_0,n} l_{m_0-1,n_0}\right)$ contains all second-order information for $m_0$. Therefore, the problems in Equations (36) and (35) have the same projection expansions, and the computation of projection expansions at level $m_0-1$ has been completed.

Note that solving the projection expansions of $m_0-1$ only requires the projection functions of $m_0$ and the optimal solution of $m_0-1$; obtaining the optimal solution of $m_0-1$ only requires solving the projection expansions of level $m_0$. To summarize, solving the optimal solution and projection expansions of any level only requires solving the optimal solution and projection expansions of its lower level. Hence, a multilevel decomposition method for the coordination of any number of levels has been realized.

The full implementation for multilevel decomposition coordination is given below. Two functions are defined first.

**Function "compute_projection"**

Input: Optimization problem formulation, upper boundary values.

Output: First- and second-order Taylor series expansions of projection functions, expanded for the input boundary value.

The implementation of this function strictly follows the procedures in Sub-section 3.2 Lower Problem Projection by replacing Equation (7) with the formulation of the input optimization problem. Hence, the implementation is not repeated here.

**Function "compute_optimum"**

Inputs: Problem number, $n^*$, level number, $m^*$, upper boundary values, $I_{m^*,n^*}\tilde{l}_{m^*-1,U(m^*,n^*)}$.

Output: Optimal solution, formulation of the converged upper problem.

For a multilevel coordination problem with $m_0$ levels, this function computes the optimal solution to the coordination problem of the sub-problem $n^*$ in $m^*$, along with all lower problems that it includes:

$$\min_{x_{m^*,n^*},u_{m^*,n^*},l_{m^*,n^*},x_{m,n},u_{m,n},l_{m,n}} f_{m^*,n^*}\left(x_{m^*,n^*}\right) + \sum_{(m,n)\in T(m^*,n^*)} f_{m,n}\left(x_{m,n}\right)$$

$$\text{s.t. } G_{m^*,n^*}\left(x_{m^*,n^*},u_{m^*,n^*},l_{m^*,n^*}\right) \leq 0$$

$$u_{m^*,n^*} = I_{m^*,n^*}\tilde{l}_{m^*-1,U(m^*,n^*)} \qquad (37)$$

$$G_{m,n}\left(x_{m,n},u_{m,n},l_{m,n}\right) \leq 0, \forall (m,n) \in T(m^*,n^*)$$

$$u_{m,n} = I_{m,n}l_{m-1,U(m,n)}, \forall (m,n) \in T(m^*,n^*)$$

where $T(m^*,n^*)$ is the set of all problems under the tree structure with $n^*$ in $m^*$ at its apex. In the above problem, if $m^* = 1$, then $u_{m^*,n^*}$ is an empty vector; furthermore, if $m = m_0$ or $m^* = m_0$, then $l_{m,n}$ or $l_{m^*,n^*}$ is an empty vector.

The implementation of this function has a decomposed structure. The original multilevel optimization problem is a special case with $m^* = 1$ in Equation (37).

The implementation of function "compute_optimum" can be summarized as follows.

**Step 1:** if $m^* = m_0$, there is only one level in Equation (37); thus, it can be directly optimized. The converged upper problem is that equation itself. Otherwise, the iteration number, $k_{m^*,n^*} = 1$, is initialized. The following initialization problem is solved at $m^*$:

$$\min_{x_{m^*,n^*}, u_{m^*,n^*}, l_{m^*,n^*}} f_{m^*,n^*}\left(x_{m^*,n^*}\right)$$

$$\text{s.t. } G_{m^*,n^*}\left(x_{m^*,n^*}, u_{m^*,n^*}, l_{m^*,n^*}\right) \leq 0 \tag{38}$$

$$u_{m^*,n^*} = I_{m^*,n^*} \tilde{l}_{m^*-1, U(m^*,n^*)}$$

The optimal solution of $l_{m^*,n^*}$ in Equation (38) is denoted $l_{m^*,n^*}^{(k)}$. $I_{m^*+1,n} l_{m^*,n^*}^{(k)}$ is sent to each adjacent lower problem, $n$, in $m^*+1$.

**Step 2:** Each adjacent lower problem calls the function "compute_optimum" with inputs: $n$, $m^*+1$, and the upper boundary values, $I_{m^*+1,n} l_{m^*,n^*}^{(k)}$. The formulation of the converged upper problem is achieved using the outputs from "compute_optimum." The function "compute_projection" is then called with the following inputs: the converged upper problem formulation and $I_{m^*+1,n} l_{m^*,n^*}^{(k)}$. The first- and second-order Taylor series expansions of the projection function are achieved with the outputs from "compute_projection," denoted $JF_{m^*+1,n}^{(k_{m^*+1,n})}\left(I_{m^*+1,n} l_{m^*,n^*}\right)$ and $JS_{m^*+1,n}^{(k_{m^*+1,n})}\left(I_{m^*+1,n} l_{m^*,n^*}\right)$, respectively. These two expansions are then sent to $n^*$ in $m^*$.

**Step 3:** $n^*$ in $m^*$ coordinates the sub-problems by optimizing the following problem:

$$\min_{x_{m^*,n^*}, u_{m^*,n^*}, l_{m^*,n^*}, o_{m^*+1,n}} f_{m^*,n^*}\left(x_{m^*,n^*}\right) + \sum_{n \in L(m^*,n^*)} o_{m^*+1,n}$$

$$\text{s.t. } G_{m^*,n^*}\left(x_{m^*,n^*}, u_{m^*,n^*}, l_{m^*,n^*}\right) \leq 0$$

$$u_{m^*,n^*} = I_{m^*,n^*} \tilde{l}_{m^*-1, U(m^*,n^*)} \tag{39}$$

$$o_{m^*+1,n} \geq JS_{m^*+1,n}^{(k_{m^*,n^*})}\left(I_{m^*+1,n} l_{m^*,n^*}\right), \forall n \in L(m^*,n^*)$$

$$o_{m^*+1,n} \geq JF_{m^*+1,n}^{(i)}\left(I_{m^*+1,n} l_{m^*,n^*}\right), \forall i = 1,2,\ldots,k_{m^*,n^*}-1, \forall n \in L(m^*,n^*)$$

$k_{m^*,n^*}$ is iterated by 1. The optimal solutions for Equation (39) are denoted $x_{m^*,n^*}^{(k_{m^*,n^*})}$, $u_{m^*,n^*}^{(k_{m^*,n^*})}$, $l_{m^*,n^*}^{(k_{m^*,n^*})}$, and $o_{m^*+1,n}^{(k_{m^*,n^*})}$.

**Step 4:** Convergence is checked. The convergence criterion is $\left\| l_{m^*,n^*}^{(k_{m^*,n^*})} - l_{m^*,n^*}^{(k_{m^*,n^*}-1)} \right\|_2 \leq \varepsilon$. If converged, $x_{m^*,n^*}^{(k_{m^*,n^*})}$, $u_{m^*,n^*}^{(k_{m^*,n^*})}$, and $l_{m^*,n^*}^{(k_{m^*,n^*})}$ are output as the optimal solutions and Equation (39) is output as the converged upper problem formulation; otherwise, the procedure returns to step 2.

## 5. Convergence and Optimality

We shall first prove that the convergence and optimality of the multilevel nested decomposition method are necessary for the convergence and optimality of the bilevel decomposition method.

**Proposition 1:** For convex optimization problems with continuously differentiable objectives and constraints, if the proposed decomposition method can converge to the global optimum of any bilevel coordination, the proposed nested decomposition method can converge to the global optimum of any multilevel coordination.

**Proof:** We assume that the nested decomposition method for the coordination of problems with $m_0$ levels has guaranteed convergence and optimality. The nested decomposition method for the coordination of problems with $m_0 + 1$ levels can be decomposed into an upper problem (i.e., the top level) and several lower problems (all problems with $m_0$ levels). Each lower problem's convergence and optimality are guaranteed, as the above decomposition of the problem with $m_0 + 1$ levels is a bilevel decomposition method; thus, the convergence and optimality of the nested decomposition method for the coordination of problems with $m_0 + 1$ levels are guaranteed. Because the multilevel nested decomposition method falls back to bilevel decomposition method when $m_0 = 2$, the proposition is proven.

∎

Next, we focus on the convergence and optimality of the bilevel decomposition method. However, a rigorous proof of the bilevel decomposition method requires modification of the iteration procedure. Specifically, when the upper problem optimizes Equation (34), it simultaneously optimizes the following problem without second-order Taylor series expansions of the lower problem projections:

$$\min_{x_1, l_1, o_{2,n}} f_1(x_1) + \sum_{n \in N(2)} o_{2,n}$$
$$\text{s.t.} \quad G_1(x_1, l_1) \leq 0 \tag{40}$$
$$o_{2,n} \geq JF_{2,n}^{(i)}(I_{2,n}l_1), \forall i = 1, 2, ..., k, \forall n \in N(2)$$

The optimal solutions of Equation (34) are denoted $x_1^{(k)}$, $l_1^{(k)}$, and $o_{2,n}^{(k)}$, while those of Equation (40) are denoted $\tilde{x}_1^{(k)}$, $\tilde{l}_1^{(k)}$, and $\tilde{o}_{2,n}^{(k)}$ (after $k$ has iterated in step 4).

If the iteration does not converge in step 5, the procedure determines whether $\tilde{x}_1^{(k)}$, $\tilde{l}_1^{(k)}$, and $\tilde{o}_{2,n}^{(k)}$ are different from $\tilde{x}_1^{(k-1)}$, $\tilde{l}_1^{(k-1)}$, and $\tilde{o}_{2,n}^{(k-1)}$. If differences exist, the original procedure is unchanged and thus returns to step 2; otherwise, the procedure returns to iteration $k-1$, sends the boundary parameter value $I_{2,n}\tilde{l}_1^{(k-1)}$ instead of $I_{2,n}l_1^{(k-1)}$ to each lower problem, and returns to step 2.

This modified procedure guarantees that the solutions of adjacent iterations of $\tilde{x}_1^{(k)}$, $\tilde{l}_1^{(k)}$, and $\tilde{o}_{2,n}^{(k)}$ from Equation

(40) are different, unless they have converged.

**Proposition 2:** For convex optimization problems with continuously differentiable objectives and constraints, the proposed bilevel decomposition method has finite convergence.

**Proof:** Consider solutions $\tilde{x}_1^{(k)}$, $\tilde{l}_1^{(k)}$, and $\tilde{o}_{2,n}^{(k)}$ of Equation (40), which are not optimal. The objective $f_1(\tilde{x}_1^{(k)}) + \sum_{n \in N(2)} \tilde{o}_{2,n}^{(k)}$ is no less than the objective $f_1(\tilde{x}_1^{(k-1)}) + \sum_{n \in N(2)} \tilde{o}_{2,n}^{(k-1)}$ from the previous iteration. Therefore, the sequence $\left\{ f_1(\tilde{x}_1^{(k)}) + \sum_{n \in N(2)} \tilde{o}_{2,n}^{(k)} \right\}$ does not decrease. Note that this sequence is also upper bounded because, for each previous iteration, the globally optimal solution of Equation (40) was a possible solution of that iteration; however, the solution was sub-optimal within the previous iteration, which means that the objective value of Equation (40) at the globally optimal solution is an upper bound of the sequence $\left\{ f_1(\tilde{x}_1^{(k)}) + \sum_{n \in N(2)} \tilde{o}_{2,n}^{(k)} \right\}$. Therefore, convergence of the iteration procedure is guaranteed according to the monotone convergence theorem.

∎

**Proposition 3:** For convex optimization problems with continuously differentiable objectives and constraints, the converged solution of the proposed bilevel decomposition method is the globally optimal solution.

**Proof:** The optimality of the converged solution can be proven by assuming the opposite. If the converged solution is not the globally optimal solution, the direction from the converged solution to the globally optimal solution is the direction that reduces the objective. Because the upper coordination problem is second-order exact, that direction also reduces the objective of the upper coordination problem. This contradicts the assumption that the solution converged. Therefore, the optimality of the bilevel decomposition method is also proven.

∎

We did not mention the modifications in this section when introducing the method in Section 3, because that method is practical and very straightforward. Moreover, the method without modifications successfully converged for each case that we tested, because the second-order Taylor series expansions of the projection functions approximate the original projection functions with exceedingly small errors around the expansion point. Therefore, these modifications typically do not have significant effects regarding the iteration procedure (i.e., the determination of differences remains intact). Therefore, the method as presented in Section 3 was not rigorous, although it avoided unnecessary complications.

## 6. Numerical Simulation Analysis

We tested the proposed nested decomposition method with several case studies. The first case study involved small-scale mathematical programming; the second case study applied the proposed method to the decomposed economic dispatch of a trilevel hierarchical power grid; the third case study applied the proposed method to the decomposed optimal power flow of a trilevel hierarchical power grid.

**6.1 Case I – Mathematical Programming**

To test the validity and efficiency of the proposed decomposition method, the following trilevel coordinated mathematical programming was tested:

$$\min_{x, y_2, y_2, z} (x-1)^2 + (y_1-2)^2 + (z-2)^2$$
$$\text{s.t. } y_1^2 + y_2^2 \leq x^2$$
$$z^2 \leq y_2^2 \tag{41}$$
$$x \geq 0$$
$$y_2 \geq 0$$

This problem has a second-order objective function and second-order cone constraints. The variable $x$ is considered the first-level variable; $y_1$ and $y_2$ are considered second-level variables; and $z$ is considered the third-level variable. Therefore, the above problem is a trilevel coordinated optimization problem, where the first-level problem is as follows:

$$\min_x (x-1)^2$$
$$\text{s.t. } x \geq 0 \tag{42}$$

The second-level problem is as follows:

$$\min_{y_2, y_2} (y_1-2)^2$$
$$\text{s.t. } y_1^2 + y_2^2 \leq x^2 \tag{43}$$
$$y_2 \geq 0$$

The third-level problem is as follows:

$$\min_z (z-2)^2$$
$$\text{s.t. } z^2 \leq y_2^2 \tag{44}$$

This trilevel coordinated optimization problem was solved by the following methods: (1) the proposed method; (2) Benders decomposition; and (3) ADMM. Because ADMM has a hyperparameter, the augmented Lagrangian parameter, $\rho$, we tested different values of this parameter.

Table I lists the results; the inner loop iteration number is the sum for all inner loops across all outer loops. The proposed method exhibits considerable advantages in terms of the number of iterations and computational efficiency.

Table I. Comparison of iteration number and time consumption among methods and among convergence criterion accuracies. ADMM: alternating direction method of multipliers.

| | Convergence criterion accuracy | $1 \times 10^{-4}$ | $1 \times 10^{-5}$ | $1 \times 10^{-6}$ |
|---|---|---|---|---|
| Proposed method | Outer loop iteration number | 2 | 2 | 2 |
| | Inner loop iteration number | 4 | 4 | 4 |
| | Total time required (s) | 0.23 | 0.23 | 0.29 |
| Benders decomposition | Outer loop iteration number | 3 | 10 | 13 |
| | Inner loop iteration number | 31 | 140 | 226 |
| | Total time required (s) | 1.42 | 5.84 | 9.75 |
| ADMM ($\rho = 3$) | Outer loop iteration number | 15 | 19 | 22 |
| | Inner loop iteration number | 211 | 326 | 463 |
| | Total time required (s) | 9.24 | 16.05 | 21.69 |
| ADMM ($\rho = 10$) | Outer loop iteration number | 28 | 35 | 42 |
| | Inner loop iteration number | 657 | 1035 | 1494 |
| | Total time required (s) | 29.80 | 54.03 | 89.16 |

**6.2 Case II – Dynamic Economic Dispatch**

The proposed method was tested on the dynamic economic dispatch of a trilevel electrical power grid, which contains a 14-bus transmission grid in the top level, three 69-bus distribution grids in the middle level, and nine 4-bus microgrids in the bottom level. The distribution grids were connected to buses 10, 11, and 12 of the transmission grid; the microgrids were connected to buses 27, 35, and 46 of each distribution grid.

The dynamic economic dispatch of an electrical power grid optimizes its total generation cost while satisfying operational and security constraints. It can be written as quadratic programming; we introduce our model in the Appendix. Day-ahead economic dispatch with 24 dispatch periods in total was considered. Thus, the boundary variables between the upper and lower power grids in this case have 24 dimensions.

The proposed decomposition method was compared with a centralized method and an isolated method. In the centralized method, the coordinated model of the trilevel electrical power grid was solved directly; thus, it was guaranteed to output the correct and accurate result. However, this centralized method cannot be implemented practically considering the operational independence and information privacy of each local of the electrical power grid. In the isolated method, boundary values between adjacent levels were negotiated in advance; the optimization of each local power grid was then conducted with fixed boundary values. The resulting generation costs of the different parts of the power grid are listed in Table II.

Table II. Comparison of total generation costs among methods

|  | Proposed decomposition method | Centralized method | Isolated method |
|---|---|---|---|
| Transmission grid | $201,224 | $201,224 | $231,096 |
| Distribution grid | $120,146 | $120,146 | $99,246 |
| Microgrid | $17,711 | $17,711 | $9,547 |
| Total | $339,081 | $339,081 | $339,889 |

The proposed decomposition method required 6 iterations for the outer loop to converge and an average of 5.2 iterations for the inner loop to converge. Comparison of the results (Table II) revealed that the proposed decomposition method achieved the same solution as the centralized method, thus demonstrating the exactness of the proposed method. Moreover, comparison with the isolated method showed that improved coordination across the trilevel power grid reduced the total generation cost by allocating generation resources among different power grids in a more economical manner.

We compared the coordination performance of the proposed decomposition method with the performances of Benders decomposition and ADMM. Notably, Benders decomposition failed to converge within 5000 iterations for the coordination between distribution grids and microgrids, because the boundary variables exhibited excessive numbers of dimensions for coordination. The iteration numbers of the proposed method and ADMM are listed in Table III.

Table III. Comparison of iteration numbers among methods. ADMM: alternating direction method of multipliers.

|  | Proposed method | ADMM | Benders decomposition |
|---|---|---|---|
| Transmission & distribution grids | 6 | 64 |  |
| Distribution grid 1 & microgrids | 28 | 2073 | Did not converge within 5000 iterations |
| Distribution grid 2 & microgrids | 27 | 2075 |  |
| Distribution grid 3 & microgrids | 38 | 2075 |  |
| Total iteration number | 99 | 6287 |  |

Convergence occurred in 108.1 and 535.6 s for the proposed method and ADMM, respectively, which indicated that the proposed method converged much more rapidly than the existing methods and proved that the proposed method is a practical improvement.

### 6.3 Case III – Optimal Power Flow

A similar, trilevel, electrical power grid was used to test the proposed decomposition method for optimal power flow. The flow model differs from the model of dynamic economic dispatch in that it uses a more accurate, nonlinear approach for its operational constraints. Specifically, the optimal power-flow models of these distribution grids and microgrids demonstrated second-order cone programming and were thus convex. Accordingly, the optimal power flow was more complex, compared with dynamic economic dispatch. The particular formulations of the optimal power flow are given in the Appendix. The boundary variables between the upper and lower power grids have three dimensions: boundary active power flow, boundary reactive power flow, and boundary voltage magnitude.

As before, the proposed decomposition method was compared with the centralized and isolated methods among power grids at each level. The total generation costs of the different methods are listed in Table IV.

Table IV. Comparison of total generation costs among methods

|  | Proposed decomposition method | Centralized method | Isolated method |
|---|---|---|---|
| Transmission grid | $11,326 | $11,326 | $12,570 |
| Distribution grid | $6,406 | $6,406 | $5,546 |
| Microgrid | $846 | $846 | $514 |
| Total | $18,578 | $18,578 | $18,630 |

The proposed decomposition method required 7 iterations for the outer loop to converge and an average of 4.8 iterations for the inner loop to converge. Note that the identical results for the proposed and centralized methods indicate the exactness and validity of the proposed method. There is an obvious economic benefit in the reduced total generation for the proposed method compared to the isolated method, thereby demonstrating the necessity of coordination among electrical power grid levels.

The proposed decomposition method was also compared with Benders decomposition and ADMM. In contrast to dynamic economic dispatch, ADMM failed to converge within 5000 iterations because of the nonlinearity of the optimal power-flow problem. Hence, the proposed method was the only method that stably converged for both the dynamic economic dispatch and optimal power-flow computation. Iteration numbers for the proposed method and Benders decomposition are listed in Table V.

Table V. Comparison of iteration numbers among methods. ADMM: alternating direction method of multipliers.

|  | Proposed method | ADMM | Benders decomposition |
|---|---|---|---|
| Transmission & distribution grids | 7 |  | 18 |
| Distribution grid 1 & microgrids | 33 | Did not converge within 5000 iterations | 160 |
| Distribution grid 2 & microgrids | 33 |  | 163 |
| Distribution grid 3 & microgrids | 34 |  | 141 |
| Total iteration number | 107 |  | 482 |

The proposed method and Benders decomposition converged in 37.3 and 123.5 s, respectively. The proposed method reduced the iteration number and computation time, thereby gaining considerable coordination efficiency by introducing second-order information for each iteration. Hence, the proposed method is more practical for reducing the communication burden between upper- and lower-level local power grids.

## 7. Conclusions

Electrical power grids are multilevel, multiarea, and hierarchically interconnected. Considering the operational independence and information privacy of each lower power grid, this paper proposed a nested decomposition method for the coordinated operation of multilevel, hierarchical, electrical power grids. A bilevel decomposition method was developed by first iterating between the upper-level problem and each lower-level problem. During each iteration, values of boundary coupling variables were computed for the upper-level problem and sent to each lower-level problem, for which an approximate projection function was computed in the boundary variable space. Because second-order information of both nonlinear objective and nonlinear constraints was considered, the proposed bilevel decomposition method has substantially better convergence than existing decomposition methods. The nested decomposition framework was extended from the proposed bilevel coordination into multilevel coordination. Our analysis showed that the converged upper-level coordination solution contained the full second-order information of the related global problem; thus, it could be used to compute approximate projection functions. The convergence and optimality of the proposed nested decomposition method were proven theoretically. We demonstrated the validity, exactness, and performance of the proposed method using several case studies regarding mathematical programming problems and multilevel power grid operations, which included both linear and nonlinear constraints.

## Acknowledgments

This work was supported by the National Natural Science Foundation of China (grant number 51725703).

## Appendix

### A. Formulation of Dynamic Economic Dispatch

The dynamic economic dispatch of a power grid optimizes the total generation cost among all dispatch periods:

$$\min \sum_{t \in T} \sum_{i \in G} C_i \left( P_{i,t}^G \right) \tag{45}$$

where $T$ is the set of dispatch periods, $G$ is the set of generator indices, $P_{i,t}^G$ is the active power output of generator $i$ at dispatch period $t$, and the function $C_i \left( P_{i,t}^G \right)$ is the generation cost function of $i$, which can be written as the following convex quadratic function:

$$C_i \left( P_{i,t}^G \right) = a_{0,i} + a_{1,i} P_{i,t}^G + a_{2,i} \left( P_{i,t}^G \right)^2 \tag{46}$$

where $a_{0,i}$, $a_{1,i}$, and $a_{2,i}$ are the constant, linear, and quadratic generation cost coefficients, respectively, of $i$.

In the linear dispatch model, constraints include the following:

(a) power balance constraints,

$$\sum_{i \in G} P_{i,t}^G = \sum_{i \in D} P_{i,t}^D, \forall t \in T \quad (47)$$

where $D$ is the set of load indices and $P_{i,t}^D$ is the active power demand of load $i$ at dispatch period $t$.

(b) generator output limits,

$$\underline{P_i^G} \leq P_{i,t}^G \leq \overline{P_i^G}, \forall i \in G, \forall t \in T \quad (48)$$

where $\overline{P_i^G}$ and $\underline{P_i^G}$ are the active power output upper and lower bounds, respectively, of generator $i$.

(c) ramping constraints,

$$-R_i^D \Delta T \leq P_{i,t}^G - P_{i,t-1}^G \leq R_i^U \Delta T, \forall i \in G, \forall t \in T \setminus \{1\} \quad (49)$$

where $R_i^U$ and $R_i^D$ are the upward and downward ramping capabilities, respectively, of generator $i$; $\Delta T$ is the dispatch interval.

(d) spinning reserve constraints,

$$0 \leq P_{i,t}^{GU} \leq R_i^U \Delta T, P_{i,t}^{GU} \leq \overline{P_i^G} - P_{i,t}^G, \forall i \in G, \forall t \in T \quad (50)$$

$$0 \leq P_{i,t}^{GD} \leq R_i^D \Delta T, P_{i,t}^{GD} \leq P_{i,t}^G - \underline{P_i^G}, \forall i \in G, \forall t \in T \quad (51)$$

$$\sum_{i \in G} P_{i,t}^{GU} \geq S_t^U, \sum_{i \in G} P_{i,t}^{GD} \geq S_t^D, \forall t \in T \quad (52)$$

where $P_{i,t}^{GU}$ and $P_{i,t}^{GD}$ are the upward and downward spinning reserve contributions, respectively, of generator $i$ at dispatch period $t$; $S_t^U$ and $S_t^D$ are the upward and downward spinning reserve requirements, respectively, of the power grid at $t$.

(e) transmission line capacity limits,

$$-P_j^L \leq \sum_{i \in G} F_{j-i}^G P_{i,t}^G - \sum_{i \in D} F_{j-i}^D P_{i,t}^D \leq P_j^L, \forall j \in L, \forall t \in T \quad (53)$$

where $P_j^L$ is the transmission capacity of line $j$, $F_{j-i}^G$ is the power transfer distribution factor from generator $i$ to line $j$, $F_{j-i}^D$ is the power transfer distribution factor from $i$ to $j$, and $L$ is the set of line indices.

When power grids are multilevel, multiarea, and hierarchically interconnected, each externally connected power grid can be considered a virtual generator from the local power grid. The boundary constraints can be modeled as the match of active power output between virtual generators.

In the above dynamic economic dispatch model, the decision variables include $P_{i,t}^G$, $P_{i,t}^{GU}$, and $P_{i,t}^{GD}$; the remaining variables are parameters. The above model has quadratic objective and linear constraints, and thus requires quadratic programming.

**B. Formulation of Optimal Power Flow**

The optimal power flow only considers one dispatch period, in which the objective is minimizing the total generation cost:

$$\min \sum_{i \in G} C_i\left(P_i^G\right) \tag{54}$$

where $G$ is the set of generator indices, $P_i^G$ is the active power output of generator $i$, and the function $C_i\left(P_i^G\right)$ is the generation cost function of $i$, which can be written as the following convex quadratic function:

$$C_i\left(P_i^G\right) = a_{0,i} + a_{1,i} P_i^G + a_{2,i}\left(P_i^G\right)^2 \tag{55}$$

where $a_{0,i}$, $a_{1,i}$, and $a_{2,i}$ are the constant, linear, and quadratic generation cost coefficients, respectively, of $i$.

Depending on whether the local power grid exhibits a mesh or radial form, different models can be established. For meshed power grids, constraints include the following:

(a) AC power-flow constraints,

$$P_{ij} = \frac{1}{\tau_{ij}^2} g_{ij}^\varepsilon V_i^2 - \frac{1}{\tau_{ij}} V_i V_j \left[ g_{ij}^\varepsilon \cos\left(\theta_i - \theta_j - \phi_{ij}\right) + b_{ij}^\varepsilon \sin\left(\theta_i - \theta_j - \phi_{ij}\right) \right], \forall ij \in L \tag{56}$$

$$P_{ji} = g_{ij}^\varepsilon V_j^2 - \frac{1}{\tau_{ij}} V_i V_j \left[ g_{ij}^\varepsilon \cos\left(\theta_j - \theta_i + \phi_{ij}\right) + b_{ij}^\varepsilon \sin\left(\theta_j - \theta_i + \phi_{ij}\right) \right], \forall ij \in L \tag{57}$$

$$Q_{ij} = -\frac{1}{\tau_{ij}^2}\left(b_{ij}^\varepsilon + \frac{b_{ij}^C}{2}\right) V_i^2 - \frac{1}{\tau_{ij}} V_i V_j \left[ g_{ij}^\varepsilon \sin\left(\theta_i - \theta_j - \phi_{ij}\right) - b_{ij}^\varepsilon \cos\left(\theta_i - \theta_j - \phi_{ij}\right) \right], \forall ij \in L \tag{58}$$

$$Q_{ji} = -\left(b_{ij}^\varepsilon + \frac{b_{ij}^C}{2}\right) V_j^2 - \frac{1}{\tau_{ij}} V_i V_j \left[ g_{ij}^\varepsilon \sin\left(\theta_j - \theta_i + \phi_{ij}\right) - b_{ij}^\varepsilon \cos\left(\theta_j - \theta_i + \phi_{ij}\right) \right], \forall ij \in L \tag{59}$$

where $P_{ij}$ and $Q_{ij}$ are the active and reactive power flows, respectively, from bus $i$ to bus $j$; $P_{ji}$ and $Q_{ji}$ are the active and reactive power flows, respectively, from $j$ to $i$; $\tau_{ij}$ is the transformer tap ratio of the line $ij$; $g_{ij}^\varepsilon$ and $b_{ij}^\varepsilon$ are the conductance and susceptance, respectively, of line $ij$; $b_{ij}^C$ is the line charging susceptance of line $ij$; $V_i$ and $V_j$ are voltage magnitudes of $i$ and $j$, respectively; $\theta_i$ and $\theta_j$ are the phase angles of $i$ and $j$, respectively; $\phi_{ij}$ is the transformer phase shift angle of line, $ij$; and $L$ is the set of line indices.

(b) power balance constraints,

$$\sum_{m \in G_i} P_m^G - \sum_{j: ji \in L} P_{ij} - \sum_{j: ij \in L} P_{ij} - \sum_{m \in D_i} P_m^D - V_i^2 g_i^s = 0, \forall i \in B \tag{60}$$

$$\sum_{m \in G_i} Q_m^G - \sum_{j: ji \in L} Q_{ij} - \sum_{j: ij \in L} Q_{ij} - \sum_{m \in D_i} Q_m^D + V_i^2 b_i^s = 0, \forall i \in B \tag{61}$$

where $G_i$ and $D_i$ are the set of generator and load indices, respectively, connected to bus $i$; $P_m^G$ and $Q_m^G$ are the active and reactive power outputs, respectively, of generator $m$; $P_m^D$ and $Q_m^D$ are the active and reactive power demands, respectively, of load $m$; $g_i^s$ and $b_i^s$ are the shunt conductance and susceptance, respectively, of $i$; and $B$ is the set of bus indices.

(c) voltage magnitude limits,

$$\underline{V_i} \leq V_i \leq \overline{V_i}, \forall i \in B \tag{62}$$

where $\overline{V_i}$ and $\underline{V_i}$ are the voltage magnitude upper and lower bounds, respectively, of bus $i$.

(d) generator output limits,

$$\underline{P_i^G} \leq P_i^G \leq \overline{P_i^G}, \underline{Q_i^G} \leq Q_i^G \leq \overline{Q_i^G}, \forall i \in G \tag{63}$$

where $\overline{P_i^G}$ and $\underline{P_i^G}$ are the active power output upper and lower bounds, respectively, of generator $i$; $\overline{Q_i^G}$ and $\underline{Q_i^G}$ are the reactive power output upper and lower bounds, respectively, of $i$.

e) transmission line capacity limits,

$$P_{ij}^2 + Q_{ij}^2 \leq \overline{S_{ij}}^2, P_{ji}^2 + Q_{ji}^2 \leq \overline{S_{ij}}^2, \forall ij \in L \tag{64}$$

where $\overline{S_{ij}}$ is the capacity of line $ij$.

In the above optimal power-flow model for meshed power grids, the decision variables include $P_{ij}$, $Q_{ij}$, $P_{ji}$, $Q_{ji}$, $V_i$, $\theta_i$, $P_i^G$, and $Q_i^G$; the remaining variables are parameters. The above model has nonlinear and nonconvex constraints, and thus requires nonconvex programming.

For radial power grids, the above model remains applicable. However, a much simpler branch-optimal power-flow model (Baran and Wu 1989; Farivar and Low 2013) can also be established. Constraints of the branch optimal power-flow model include the following:

(a) branch flow equations,

$$P_{ij}^2 + Q_{ij}^2 = v_i l_{ij}, \forall ij \in L \tag{65}$$

where $P_{ij}$ and $Q_{ij}$ are the active and reactive power flows, respectively, from bus $i$ to bus $j$; $v_i$ is the square of the voltage magnitude of $i$; $l_{ij}$ is the square of the current magnitude of line $ij$; and $L$ is the set of line indices.

(b) power balance constraints,

$$\sum_{m \in G_i} P_m^G + \sum_{j:ji \in L} (P_{ji} - l_{ji} r_{ji}) = \sum_{j:ij \in L} P_{ij} + \sum_{m \in D_i} P_m^D + v_i g_i^s, \forall i \in B \tag{66}$$

$$\sum_{m \in G_i} Q_m^G + \sum_{j:ji \in L} (Q_{ji} - l_{ji} x_{ji}) = \sum_{j:ij \in L} Q_{ij} + \sum_{m \in D_i} Q_m^D - v_i b_i^s, \forall i \in B \tag{67}$$

where $G_i$ and $D_i$ are the set of generator and load indices, respectively, connected to bus $i$; $P_m^G$ and $Q_m^G$ are the active and reactive power outputs, respectively, of generator $m$; $r_{ji}$ and $x_{ji}$ are the resistance and reactance, respectively, of line $ji$; $P_m^D$ and $Q_m^D$ are the active and reactive power demands, respectively, of load $m$; $g_i^s$ and $b_i^s$ are the shunt conductance and susceptance, respectively, of bus $i$; and $B$ is the set of bus indices.

(c) branch voltage constraints,

$$v_j = v_i - 2(r_{ij} P_{ij} + x_{ij} Q_{ij}) + (r_{ij}^2 + x_{ij}^2) l_{ij}, \forall ij \in L \tag{68}$$

(d) voltage magnitude limits,

$$\underline{v_i} \leq v_i \leq \overline{v_i}, \forall i \in B \tag{69}$$

where $\overline{v_i}$ and $\underline{v_i}$ are the voltage magnitude square upper and lower bounds, respectively, of bus $i$.

(e) generator output limits,

$$\underline{P_i^G} \leq P_i^G \leq \overline{P_i^G}, \underline{Q_i^G} \leq Q_i^G \leq \overline{Q_i^G}, \forall i \in G \tag{70}$$

where $\overline{P_i^G}$ and $\underline{P_i^G}$ are the active power output upper and lower bounds, respectively, of generator $i$; $\overline{Q_i^G}$

and $\underline{Q}_i^G$ are the reactive power output upper and lower bounds, respectively, of $i$.

(f) transmission line capacity limits,

$$l_{ij} \leq \overline{l}_{ij}, \forall ij \in L \tag{71}$$

where $\overline{l}_{ij}$ is the current capacity square of line $ij$.

In the above model, the decision variables include $P_{ij}$, $Q_{ij}$, $v_i$, $l_{ij}$, $P_i^G$, and $Q_i^G$; the remaining variables are parameters. The only nonlinear constraint in the above model is Equation (65). Notably, that constraint can be further convexified into a second-order cone constraint:

$$P_{ij}^2 + Q_{ij}^2 \leq v_i l_{ij}, \forall ij \in L \tag{72}$$

The analysis by Li et al. (2012) shows that the above convexification will not introduce excessive error. Thus, the branch optimal power-flow model requires second-order cone programming after convexification.

Similar to the modeling of multilevel, multiarea, and hierarchically interconnected power grids in the dynamic economic dispatch, each external adjacent power grid can be considered a virtual generator in the local power grid. Boundary coupling constraints include matching the active power, reactive power, and voltage magnitude.

## References

Baran ME, Wu FF (1989) Optimal capacitor placement on radial distribution systems. *IEEE Transactions on Power Delivery*. 4(1):725–734.

Benders JF (1962) Partitioning procedures for solving mixed-variables programming problems. *Numerische Mathematik*. 4(1):238–252.

Boyd S, Parikh N, Chu E, Peleato B, Eckstein J (2011) Distributed optimization and statistical learning via the alternating direction method of multipliers. *Foundations and Trends® in Machine Learning*. 3(1):1–122.

Chen G, Yang Q (2017) An ADMM-based distributed algorithm for economic dispatch in islanded microgrids. *IEEE Transactions on Industrial Informatics*. 14(9):3892–3903.

Conejo AJ, Castillo E, Minguez R, Garcia-Bertrand R (2006) *Decomposition techniques in mathematical programming: engineering and science applications*. Springer Science & Business Media.

Farivar M, Low SH (2013) Branch flow model: relaxations and convexification—Part I. *IEEE Transactions on Power Systems*. 28(3):2554–2564.

Farivar M, Low SH (2013) Branch flow model: relaxations and convexification—Part II. *IEEE Transactions on Power Systems*. 3(28):2565–2572.

**Lin and Wu:** *Nested Decomposition Method*

27Geoffrion AM (1972) Generalized benders decomposition. *Journal of Optimization Theory and Applications*. 10(4):237–260.

Hestenes MR (1969) Multiplier and gradient methods. *Journal of Optimization Theory and Applications*. 4(5):303–320.

Li N, Chen L, Low SH (2012, November) Exact convex relaxation of OPF for radial networks using branch flow model. In *2012 IEEE Third International Conference on Smart Grid Communications (SmartGridComm)* (IEEE), 7–12.

Li Z, Wu W, Zhang B, Wang B (2015) Decentralized multi-area dynamic economic dispatch using modified generalized benders decomposition. *IEEE Transactions on Power Systems*. 31(1):526–538.

Li Z, Wu W, Zeng B, Shahidehpour M, Zhang B (2016) Decentralized contingency-constrained tie-line scheduling for multi-area power grids. *IEEE Transactions on Power Systems*. 32(1):354–367.

Lin C, Wu W, Zhang B, Wang B, Zheng W, Li Z (2016) Decentralized reactive power optimization method for transmission and distribution networks accommodating large-scale DG integration. *IEEE Transactions on Sustainable Energy*. 8(1):363–373.

Moya OE (2005) A spinning reserve, load shedding, and economic dispatch solution by Bender's decomposition. *IEEE Transactions on Power Systems*. 20(1):384–388.

Powell MJ (1969) A method for nonlinear constraints in minimization problems. *Optimization*, 283-298.

Sifuentes WS, Vargas A (2007) Hydrothermal scheduling using benders decomposition: accelerating techniques. *IEEE Transactions on Power Systems*. 22(3):1351–1359.

Wang SJ, Shahidehpour SM, Kirschen DS, Mokhtari S, Irisarri GD (1995) Short-term generation scheduling with transmission and environmental constraints using an augmented Lagrangian relaxation. *IEEE Transactions on Power Systems*. 10(3):1294–1301.

Zheng W, Wu W, Zhang B, Sun H, Liu Y (2015) A fully distributed reactive power optimization and control method for active distribution networks. *IEEE Transactions on Smart Grid*. 7(2):1021–1033.